\documentstyle[preprint,floats,aps,epsf,prb,amsmath,epsfig]{revtex}
\newcommand \ltdash{\raise-1.8pt\hbox{$\scriptscriptstyle |$}}

\newcommand \bea {\begin{eqnarray} }
\newcommand \eea {\end{eqnarray}}

\newcommand \uru {$URu_{2}Si_{2}$\ }

\newlength{\bxwidth}\bxwidth=0.8\textwidth

\newcommand\prk[2]{$ $\vskip 2.4 truein\vskip -2 truein
\centerline {\epsfig{file=#1,width=\bxwidth}} \vskip 0.5truein
\centerline{{\bf Fig.} #2}}
\begin{document}
\title{The Case for Phase Separation in $URu_2Si_2$
}
\author{P. Chandra,$^1$ P. Coleman,$^2$ J.A. Mydosh$^{3,4}$ and V. Tripathi$^5$  
}
\address
{$^{1}$ Chandra Tech Consulting LLC, 123 Harper St, Highland Park, NJ
08904, USA}
\address
{$^{2}$ Center for Materials Theory,
Rutgers University, Piscataway, NJ 08855, U.S.A. }  
\address
{$^{3}$ Max Planck Institute for Chemical Physics of Solids, 01187 Dresden, Germany}
\address
{$^{4}$Kamerlingh Onnes Laboratory,  Leiden University,
P. O. Box 9504, 2300 RA Leiden, The Netherlands}
\address
{$^{5}$ TCM Group, Cavendish Laboratory, Madingley Road, Cambridge CB3 0HE,UK
}  
\maketitle
\begin{abstract}
Motivated by experiment, we review the case for phase inhomogeneity in \uru. 
In this scenario, the paramagnetic hidden order phase coexists with small distinct
domains of antiferromagnetism whose volume fraction increases with
pressure.  The implications for the nature of the hidden order are discussed.

\end{abstract}
\eject
The heavy fermion material \uru poses a unique challenge.  Discovered
almost two decades ago, it provides a classic example of a
mean-field phase transition\cite{Chandra94} at $T_{c}=17K$; yet there is still no consensus
on the nature of the underlying order. More specifically, the transition is
characterized by sharp anomalies in a number of 
bulk properties\cite{Palstra85,Miyako91,Ramirez92} and
a gap\cite{Walter86,Mason91,Broholm87,Broholm91} that each develop at $T_c$.
Initially the ordered phase of
this material was characterized as a spin density wave,
but subsequent neutron scattering measurements\cite{Broholm87,Broholm91} indicated 
that the size of the staggered moment is too
small ($\sim 0.02 - 0.04 \mu_B$ per uranium atom) to account for the 
substantial entropy loss which occurs at the
transition.\cite{Buyers96}   

A sequence of recent experimental developments has led to 
new insight into the nature of the hidden order in \uru.
High-field measurements\cite{Mentink96,Bourdarot03} have revealed that the staggered 
magnetization and the gap have {\sl different} field-dependences,
suggesting that there are two distinct order parameters, $M$ and $\psi $.
Initial theories assumed that the hidden
order and the spin antiferromagnetism were coupled and spatially homogeneous.\cite{Shah00}  
Pressure-dependent neutron scattering studies\cite{Amitsuka99} subsequently showed that the ordered
antiferromagnetic moment $M$ in \uru grows roughly linearly with
applied pressure $M \propto P$ up to $P_0=1$ GPa.  
Within the homogeneous scenario, this result requires a
pressure-dependent coupling between the hidden order and the
magnetism ($\Delta {\cal F}\sim - \psi  M P$);
such a linear coupling, required to nearly vanish at ambient pressure,  
is awkward to justify on symmetry grounds.  

Pressure-dependent NMR studies\cite{Matsuda01} provided a natural
resolution of this dilemma by revealing that there exist
distinct antiferromagnetic and paramagnetic regions whose
relative volume fraction changes with applied pressure and temperature.
More specifically, the single resonance associated
with paramagnetism at high temperatures remains clearly at
temperatures $T < T_c$, where at finite applied pressure
it coexists with two symmetric satellite lines associated with
antiferromagnetic ordering.   
The frequency-shift associated with these additional 
resonances is independent of applied pressure, indicating
the presence of a constant magnetic moment.  However the
relative integrated intensities of the antiferromagnetic
and the paramagnetic lines are temperature- and pressure-dependent,
and are naturally interpreted as reflecting the relative volume
fractions of spin ordered and disordered regions.
Indeed, assuming a fixed magnetic moment, the pressure-dependence of
the magnetic Bragg peak observed in neutron scattering\cite{Amitsuka99}  
is consistent with the antiferromagnetic volume
fraction taken from the NMR work.\cite{Matsuda01}
The natural conclusion from these studies, supported by earlier 
$\mu$SR data,\cite{Luke94}
is that the observed increase in the magnetization as a function
of pressure is simply a volume fraction effect.\cite{Matsuda01}
Moreover these measurements indicate that at ambient
pressure there exists a large pressure-independent
moment that resides in less than 10 \% of the material.
The majority phase therefore contains no conventional spin order,
and theoretically the hidden order parameter is no longer
accountable for the small but finite presence of antiferromagnetism.

In the pressure-dependent neutron scattering experiments,\cite{Amitsuka99} 
the character of the magnetic transition changes from mean-field to Ising at
$P = P_0$.  This feature, combined with the observed linear pressure-dependence
of $M$ for $P < P_0$ is naturally interpreted as originating from
the presence of a bicritical point (Figure 1a).\cite{Chandra02}  In this scenario,
at ambient pressure the observed magnetization is a volume fraction
effect which develops {\sl distinctly} from the hidden-order
via a first-order transition.  We can study the phase behavior
of such a system using the free energy 
\begin{equation}
F = F_{\psi} + F_M + g\psi^2M^2
\end{equation}
where $F_X = (T_X(V) - T)X^2 + \frac{1}{2}u_X^2X^4$ with $X = \{\psi,M\}$
and $T_\psi = T_M$ at a critical volume $V_c$.  If $g^2 \ge u^2_{\psi} u^2_M$, a bicritical 
point exists\cite{Chaikin94} at $V=V_c$  with an associated first-order line.

Transforming the $T-V$ phase diagram 
into one for $T-P$ (Figure 1a), we remark that the pressure $P = -\frac{\partial F}{\partial V}$ 
is discontinuous across the first-order line in Fig. 1a, leading to
two {\sl distinct} pressure scales, $P_\psi$ and $P_M$ in the  $T-P$ plot
(Fig. 1b) and an associated coexistence region.  There the fraction
of the magnetic phase $x$ is given by the expression 
$P(x) = (1-x)P_{\psi} + xP_M$ so that the net magnetization is then
\begin{equation}
{\cal M} = Mx = M \left(\frac{P - P_\psi}{P_M - P_\psi}\right).
\label{M}
\end{equation} 
Equation (2) indicates the linear relation of the observed magnetization
as a function of pressure for $P > P_\psi$  where $P_\psi$ is small due to a large pressure-change 
associated with the first-order line in Fig. 1a.

There are a number of experimental observations that are consistent
with this scenario where the hidden order and the spin antiferromagnetism
are phase separated.  Within this framework the {\sl spatially} inhomogeneous
average ``moment'' is taken to be
\begin{equation}
M^2 = \frac{1}{V} \int \langle M(x)M(0)\rangle d^3 x,
\label{vmom}
\end{equation}
where $M(x)$ is the local staggered magnetization.
For a fixed site-independent value of $M(x)$, (\ref{vmom}) is 
simply proportional to the volume fraction of
antiferromagnetic regions.  Earlier $\mu$SR studies\cite{Luke94} found that the
muon precession signal, sensitive to magnetic ordering, developed
abruptly at the transition $T_c$, suggesting a first-order transition
of the magnetization. Upon cooling, this precession frequency remained
constant indicating that the size of the moment is temperature-independent.
By contrast, the amplitude of the precessing signal increased with decreasing
temperatures indicating a change in the underlying antiferromagnetic volume fraction.  
Recent $\mu$SR studies have extended this work, 
confirming the increase of the precession amplitude with applied pressure;\cite{Amitsuka03}
this result is consistent with the NMR data.\cite{Matsuda01}
At ambient pressure the onset temperatures of the hidden order and the antiferromagnetism
are very close, but they can be separated by both chemical\cite{Amitsuka03} and applied\cite{Motoyama02}
pressure. Finally such measurements\cite{Amitsuka03,Motoyama02} indicate that the onset detection of the inhomogeneous
antiferromagnetism is critically dependent on sample quality and history, particularly as a function of pressure. 
This is to be expected in a system of spatial inhomogeneities.   

An alternative proposal\cite{Bernhoeft02} to phase inhomogeneity emphasizes 
the inferred presence of a {\sl dynamical} order parameter
whose time-dependence is invoked to explain observed behavior in \uru.
In particular resonant X-ray scattering, a probe with a time-scale of 
$\tau_{\phi} \sim 10^{-14}$ seconds, indicates a moment of $0.3 \mu_B$ per 
uranium atom which is consistent with the entropy 
lost at the transition.\cite{Palstra85,Buyers96} It is 
argued that there is temporal averaging of
the moment over time-reversed Neel states
on the time-scales probed by neutron scattering
($\tau _{n}\sim 10^{-12}$ second), so that only a fraction of it
is observed ($0.02 \mu_B$ per uranium) (Figure 1b). Similarly, it is noted that 
NMR and $\mu$SR measurements on \uru, both
of which have much longer observation time-scales 
than do neutrons, indicate no long-range magnetic order at all.
In this scenario, the {\sl temporally} inhomogeneous average ``moment''
is taken to be
\begin{equation}
M^2 = \frac{1}{\tau_m} \int \langle M(\tau)M(0)\rangle d \tau
\label{tmom}
\end{equation}
where $M(\tau)$ is the dynamical staggered magnetization.  Here 
$\tau_m$ refers to the measurement time; because the moment fluctuates
between different orientational states, longer and longer time-averaging
occurs as $\tau_m$ is increased, leading to a decreasing value of $M$.
Application of pressure is argued to slow down
the moment fluctuations, hence making the full amplitude of
the magnetic order parameter
``accessible'' to neutrons.  Indeed
the observed saturated moment measured under pressure by neutron scattering 
is consistent with the value observed 
by fast resonant X-rays.\cite{Amitsuka99,Bernhoeft02}

It is instructive to compare the situation in \uru
with that of the pseudobinaries $U(Pt_{1-x}Pd_x)_3$.
For $x \le 0.01$, neutron\cite{Aeppli88,Hayden92} and
magnetic X-ray scattering\cite{Isaacs95} experiments
reveal a small moment ($0.02 \mu_B$ per uranium atom), comparable in magnitude
to that observed in analogous measurements of \uru at ambient pressure.
However NMR\cite{Tou96}and $\mu$SR experiments\cite{Reotier95,Keizer99} on these Pd-doped
$UPt_3$ materials do not detect this moment at all,
leading to the suggestion that it fluctuates on time-scales intermediate
between the observation time-windows of the two sets of probes. The development
of this small-moment antiferromagnetic state occurs via a crossover\cite{Visser00} rather
than a transition and is {\sl not} accompanied by
any thermodynamic anomalies.  By contrast, at higher dopings
($0.02 \le x \le 0.08$), both neutrons and muons
see a large moment ($\sim 0.6 \mu_B$) and there are associated 
discontinuities in bulk properties.\cite{Visser00} Here the key point is
that fluctuating moments do exist and their development is associated
with a crossover and {\sl not} a true phase transition; they can enhance
pre-existing thermodynamic anomalies\cite{Kos03} but they cannot produce
such discontinuities purely by themselves.
In contrast to the situation in $U(Pt_{1-x}Pd_x)_3$ for low doping ($x \le 0.01$),
in \uru there are dramatic thermodynamic signatures of a true phase transition
coexisting with the presence of a small static moments,\cite{Buyers96}
that {\sl cannot} be {\sl solely} explained by dynamical fluctuations.  
Similarly the pressure-independent frequency shift of 
the antiferromagnetic
resonance lines detected in NMR measurements\cite{Matsuda01}
at temperatures $T < T_0$ is difficult to reconcile with a homogeneous
dynamical moment whose fluctuating time-scale is reduced with applied pressure.

Recent high-field measurements (Figure 2) of the antiferromagnetic moment and
the magnetic excitations with neutron scattering\cite{Bourdarot03} 
also yield insight about the question of spatial inhomogeneity of
the spin magnetism. These experiments indicate that the
field-dependence of the moment has a distinctive
inflection point at 7 Tesla, and remains finite
but small up to fields of order 17 Tesla.  Such
behavior is strongly suggestive of a {\sl local} linear
coupling term between $M$ and $\psi$ of the form
\begin{equation}
\Delta {\cal F} = g\int d^3x M(x) \psi(x).
\label{lcoupling}
\end{equation}
Indeed it was shown earlier\cite{Shah00} that the
presence of such a term in the Landau-Ginzburg free energy  
leads to a field-dependence 
of the staggered magnetization, $M[h]$,  of the form
\begin{equation}
M^2[h]= M^2_0 \frac{[1-h^2]}{(1+\delta h^2)^2}
  ,
\label{Mh}
\end{equation}
where $h = \frac{H}{H_c}$ is the ratio of the external and the measured critical magnetic fields
($H_c = 40 T$) and  $\delta$ is defined through the relation $
T_m(V,h)-T = [T_m(V)-T](1 + \delta h^2)$.
We note that that this expression has an point of inflection around
the field value $H_m \sim H_c h_m$ where $h_m = \frac{1}{\sqrt{1 + 2 \delta}}$;
qualitatively this is because $M$ decreases with small $h$ but, due to
its coupling to $\psi$, 
must maintain a nonzero value up to $h = 1$(Figure 2).

At first sight, the presence of the observed
inflection point\cite{Bourdarot03}
is rather puzzling, particularly as the original phenomenology\cite{Shah00}
was developed for homogeneously coexisting magnetic and hidden order
parameters.  Certainly it appears
to confirm the existence of a linear coupling between $M$ and $\Psi$.
In a homogeneous system, such a term can only occur in the free energy if 
(i) $M$ and $\psi$ have
the same ordering wavevector
and (ii) $\psi$ breaks time-reversal symmetry.
From neutron scattering 
measurements\cite{Broholm87,Broholm91} it is known that $M$ is commensurate,
whereas there are many indications that the hidden order is not.
In particular, we expect $\psi$ to be incommensurate due to the fact that
the observed entropy loss and the accompanying gap suggest that it
results from a Fermi surface instability.  Furthermore the observed
insensitivity of the elastic response\cite{Luthi93} at $T_c$ is 
consistent with the presence of an incommensurate density wave that
couples weakly to uniform strain.  Thus it appears that
in a homogeneous system the presence of a linear coupling term
between the magnetization and the hidden order is unlikely
due to the dissimilarly of their respective wavevectors.
However, in a phase-separated scenario, it may be easier to
motivate the presence of such a coupling between
$M$ and $\psi$.  The spatial inhomogeneity of the spin order and the
distribution of antiferromagnetic domain sizes
mean that translational invariance is lost, making it possible for
a {\it local} coupling to develop between the two order parameters.  The presence 
of  stacking faults and other defects will tend to enhance the strength of this coupling. 
Indeed, in the presence of disorder 
it is difficult\cite{Nattermann98} 
to avoid a local linear coupling between random fields and a coexisting   
order parameter if such a term in the free energy is allowed by time-reversal
invariance.
 
The presence of an inflection in the field-dependent magnetization is
an indication that the hidden order parameter {\sl breaks}
time-reversal symmetry, consistent with previous NMR
measurements.\cite{Bernal01,Chandra02b} Motivated both by experiment
and by symmetry considerations, we have argued
elsewhere\cite{Mydosh03} that the two leading candidates for the
hidden order are a quadrupolar charge density wave and an orbital
antiferromagnet.  The key factor distinguishing these contenders is
the presence or absence of time-reversal breaking.  Thus the recent
high-field measurements\cite{Bourdarot03} point towards orbital
antiferromagnetism.  Naturally it would be optimal to have a direct
experimental test of this conjecture. Neutron scattering could provide
such a probe, particularly since the form factor associated with the
extended current loops of the orbital antiferromagnet is different
from that of point spins.\cite{Chandra02b} More specifically, we have
used the spatial distribution of the orbital fields consistent with
the NMR results\cite{Bernal01} to determine the positions, the
intensities and the form factor associated with the peaks anticipated
in neutron scattering measurements.  Perhaps most important, we
find\cite{Chandra02b} that the maximum scattering intensity is
predicted to lie in a ring $\vec Q = \vec Q_0 + \vec q$ of radius
$\vert \vec q \vert \sim 0.2$ centered around wavevector 
$\vec Q_0 = (0 0 1)$, where $\vec q$ lies in the a-b plane.  It should be noted that scattering in the vicinity of $\vec Q_0$ is {\sl forbidden} for
the case of ordered  spins aligned along the c-axis, for their dipole form factor is proportional to
$(\hat Q \times \vec{M})^2$, and thus vanishes for $\hat Q ||
\vec{M}$.  Hence the presence of neutron scattering at this particular
wavevector would be a ``smoking gun'' confirmation of incommensurate
orbital antiferromagnetism as the enigmatic hidden order.

In summary, motivated by experiment, we have presented
the case for phase separation of spin magnetism
and hidden order in \uru.  We argue that 
pressure-dependent neutron\cite{Amitsuka99} and nuclear
magnetic resonance\cite{Matsuda01} studies
are naturally interpreted in terms of a {\sl spatially}
inhomogeneous moment whose volume fraction increases
with applied pressure.  The alternative proposal
of a {\sl temporally} inhomogeneous moment that is
spatially homogeneous cannot, to our knowledge, account
for the marked entropy loss and the bulk discontinuities
associated with the transition at $T_c$.  Furthermore
recent observation of a marked inflection in the
field-dependence of the magnetization indicates
an underlying linear coupling between the
$M$ and $\psi$, which is difficult to understand
in a homogeneous scenario due to the disparateness
of their wavevectors.  By contrast, such a term
could be realized as a local coupling in a phase segregated system
where the microscopic 
spatial inhomogeneities of the order parameters
break translational symmetry. 
It is important
to emphasize that, even within the phase separation
scenario, such a coupling can {\sl only} exist
if the hidden order parameter breaks time-reversal
invariance.  Thus the field-dependent magnetization
studies,\cite{Bourdarot03}  
like earlier NMR measurements,\cite{Bernal01}
point towards incommensurate orbital
antiferromagnetism as a key contender for the hidden
order.  A direct test of this conjecture would
be neutron measurements at a particular wavevector
where scattering is forbidden for point spins.
We look forward with much anticipation to the results
of these measurements.

We acknowledge discussions with G. Aeppli, H. Amitsuka, K. McEuen and R. Walstedt.
This project is partially supported under grant NSF-DMR 9983156 (Coleman
and Tripathi).  Part of this work (P. Chandra) was performed at NEC Research
Institute and at the Aspen Center for Physics.

\vfill \eject


\newpage

\noindent{\bf Figure Captions}

\begin{enumerate}

\item 
\label{1}
Schematics graphically contrasting (a) temporally and (b) spatially
inhomogeneous moments that play key roles
in the dynamical moment and the phase separation
scenarios respectively of \uru.

\item 
\label{2}
A sketch of the data,\cite{Bourdarot02} showing the field variation of  the neutron scattering
intensity from the staggered moment, compared with the predictions
(\ref{Mh}) of Landau-Ginzburg theory. Dashed line shows variation of
hidden order gap with field, dotted line, the quadratic extrapolation
of the low field dependence of the moment. 
\end{enumerate}

\newpage

\begin{figure}

\bxwidth=\textwidth
\prk{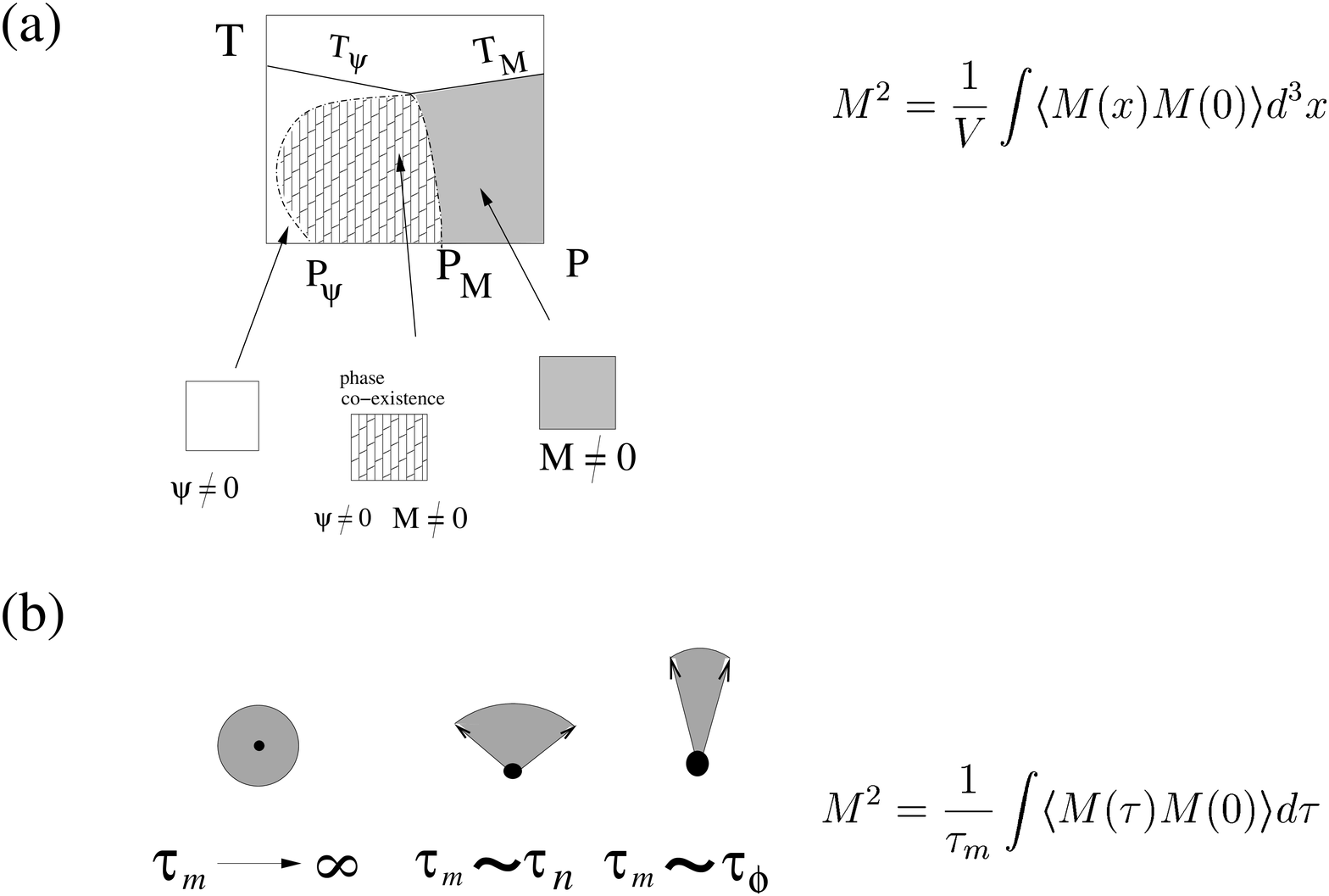}{1}

\end{figure}
\newpage

\begin{figure}

\bxwidth=\textwidth
\prk{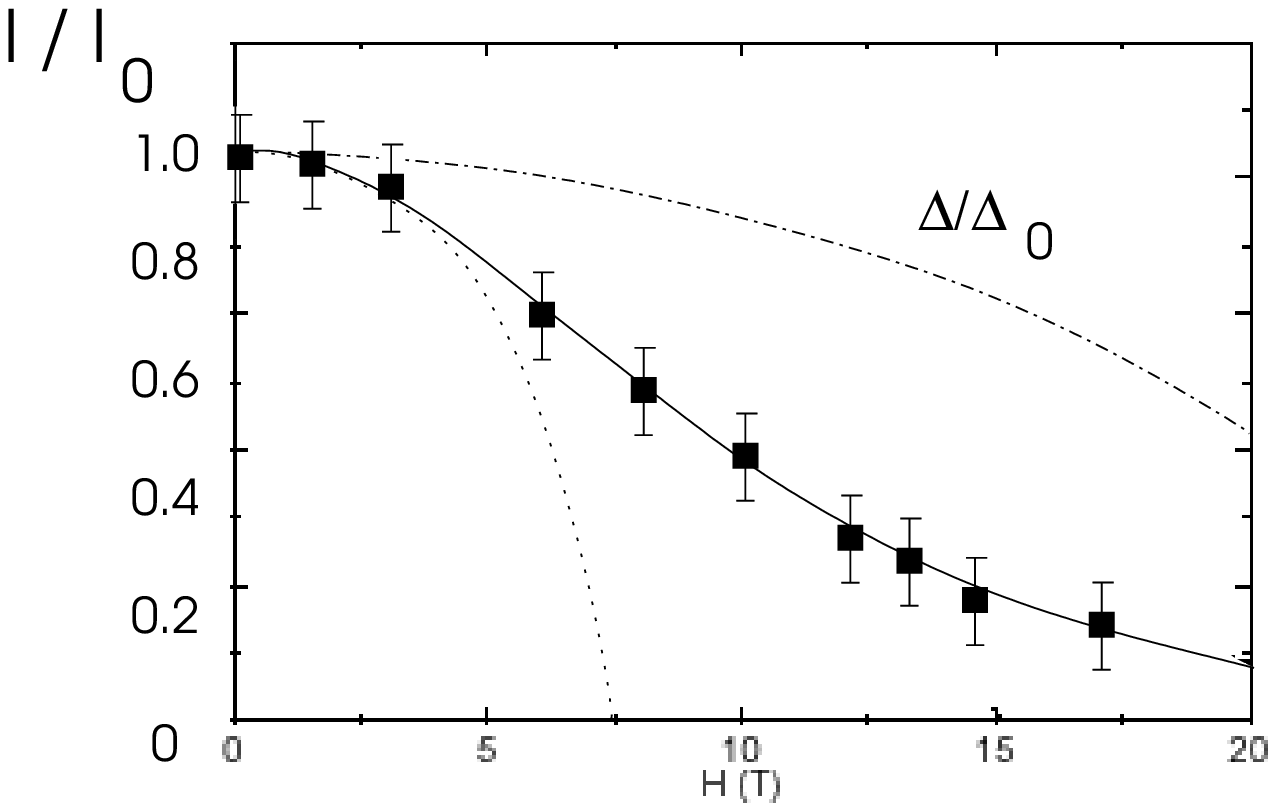}{2}

\end{figure}





\end{document}